\documentstyle[11pt,epsf]{article}

\begin{document}
%
\title{Collective Spin Modes in Superconducting Double Layers}
\author{ Christian Helm, Franz Forsthofer, and Joachim Keller\\
Institute for
    Theoretical Physics, University of  
Regensburg,\\ D-93040 Regensburg, Germany}

\date{submitted to J. of Low Temp. Phys.}

\maketitle

\begin{abstract}
We investigate 
a  double layer system  with tight-binding hopping,
intra-layer and inter-layer interactions, as well as a
Josephson like coupling.  We find that an antiferromagnetic  spin 
polarization  induces additional spin-triplet pairing (with $S_z
=0$) to 
the singlet order parameter.
This causes an undamped collective mode in the superconducting state 
below the particle-hole threshold, which is interpreted as a
Goldstone  
excitation. 
\end{abstract}


\section{INTRODUCTION AND MODEL} 

Collective density fluctuations in superconductors due to the
breakdown of  
the global gauge invariance are well known theoretically 
\cite{WuGriffin,wir}.
However, since these modes couple to charge oscillations, the long
range  
Coulomb force usually
pushes up their energies to the plasma frequency.
 One possibility  to avoid  the Coulomb interaction  completely, is  
to  consider  spin fluctuations  instead  of charge  fluctuations 
between  the layers.  
In the following   we will  show the existence of such a sharp, 
collective   spin mode in the gap, which might 
have been observed\cite{Fong} in inelastic neutron scattering on 
Y-Ba-Cu-O.

We consider an electronic  double-layer  system 
described  by the Hamiltonian 
$H=H_0+H_S$:
\begin{eqnarray}
H_0 &=& \sum_{k\sigma} \epsilon_k
 (c^\dagger_{1k\sigma}c^{\phantom{\dagger}}_{1k\sigma}
+ c^\dagger_{2k\sigma}c^{\phantom{\dagger}}_{2k\sigma})
  + t_k  (c^\dagger_{2k\sigma} 
c^{\phantom{\dagger}}_{1k\sigma}
 + c^\dagger_{1k\sigma} c^{\phantom{\dagger}}_{2k\sigma}) 
\\
H_S &=& \small{\frac{1}{2}}  \sum_{k k'q \sigma \sigma'} \sum_i
{\phantom+} 
 V_\parallel \,
 c^\dagger_{i   k+q  \sigma} 
 c^\dagger_{i  k'-q\sigma'}  c^{\phantom\dagger}_{i  k'\sigma'} 
 c^{\phantom\dagger}_{i  k  \sigma}
+V_\perp \,
 c^\dagger_{i   k+q  \sigma}  
 c^\dagger_{j  k'-q\sigma'}  c^{\phantom\dagger}_{j  k'\sigma'}
 c^{\phantom\dagger}_{i  k  \sigma} \nonumber
\\
& &{\phantom{ {1\over 2} \sum_{k k'q \sigma \sigma'}  } }
+J\,
(
 c^\dagger_{i   k+q  \sigma}  
 c^\dagger_{i  k'-q\sigma'}  c^{\phantom\dagger}_{j  k'\sigma'}
 c^{\phantom\dagger}_{j  k  \sigma}
 +c^\dagger_{i   k+q  \sigma}  
 c^\dagger_{j  k'-q\sigma'}  c^{\phantom\dagger}_{i  k'\sigma'}
 c^{\phantom\dagger}_{j  k  \sigma} 
) .
\end{eqnarray}
Here $t_k$ describes  a tight-binding  coupling  between  the two  
layers $i=(1,2), j=3-i$, while  $V_\parallel$  ($V_\perp$) are 
intra-(inter)-layer pairing interactions and the Josephson-like
coupling  
$J$ describes the coherent 
transfer  of two particles  from one layer  to the other. 
In a previous publication \cite{wir} we treated this model 
 using the Nambu formalism including vertex corrections 
to calculate charge fluctuations 
between the layers.

In this paper we are primarily  interested  in the calculation  of 
 correlation functions involving  the operator
\begin{equation}   
S = \sum_{k} 
   c^\dagger_{2k\uparrow} c^{\phantom\dagger}_{2k\uparrow} 
-c^\dagger_{2k\downarrow} c^{\phantom\dagger}_{2k\downarrow} 
-c^\dagger_{1k\uparrow} c^{\phantom\dagger}_{1k\uparrow} 
+c^\dagger_{1k\downarrow} c^{\phantom\dagger}_{1k\downarrow} 
\end{equation}
  describing  
the difference  of the spin polarization 
in the two layers  and the operators coupling to it
($\Delta_{ij}^{\dagger} := c^{\dagger}_{i k  \uparrow} 
 c^{\dagger}_{j -k \downarrow}$)
\begin{eqnarray}   
&\Phi_T =  -i \sum_{k} 
\Delta_{21}^{\dagger} - \Delta_{12}^{\dagger} - \Delta_{21} +
\Delta_{12}, 
\nonumber
\\
&M = -i \sum_{k} 
c^\dagger_{2k\uparrow} c^{\phantom\dagger}_{1k\uparrow} 
-c^\dagger_{1k\uparrow} c^{\phantom\dagger}_{2k\uparrow} 
-c^\dagger_{2k\downarrow} c^{\phantom\dagger}_{1k\downarrow} 
+c^\dagger_{1k\downarrow} c^{\phantom\dagger}_{2k\downarrow} .
&
\end{eqnarray}
The quantity $M$ corresponds  to the spin current between the two 
layers. $\Phi_T$  and $A_T$ describe  pairing in 
different layers in a spin-triplet state with total spin $S_z=0$ and
 are the real and 
imaginary  part of the inter-layer triplet-pairing amplitude
$\Delta_{\perp, T}:= \Delta_{12} - \Delta_{21}$ . 
To shorten the notation, we introduce 
\begin{equation}
P^{ij} := \sum_k \Psi_k^{\dagger} D^{ij} \Psi_k,   \,\, 
D^{ij} := \sigma_i \otimes \tau_j,  \,\,\, 
\Psi_k := ( c_{1 k \uparrow}, c_{1 -k \downarrow}^{\dagger}, c_{2 k
  \uparrow},  
          c_{2 -k \downarrow}^{\dagger} )^t 
\end{equation}
$\tau_i$ ($\sigma_i$) being the Pauli matrices in the Nambu  or
two-layer space,
respectively (examples: $S = - P^{30}, A_T = - P^{22}, \Phi_T =
-P^{21}$). 

\section{ANALYTICAL RESULTS AND  GOLDSTONE MODES}

In general  the correlation functions $\ll P^{ij},P^{lm}\gg$    have
to be determined numerically. However, 
for constant hopping $t_k = t$ with
$t,   \omega  \ll \Delta$ ($\Delta$ is the  superconducting s-wave
gap) 
and weak coupling the collective modes can be calculated
analytically (for $\omega_S, \omega_0 \ll \Delta$)
in the  cases i (ii) of  {\it pure} intra-(inter)-layer pairing.
\begin{equation}\label{correl}
\renewcommand{\arraystretch}{1.7}
\begin{array}{cccc}
\mbox{We obtain}&\mbox{for} &\mbox{case (i)}& \mbox{case (ii)}\\
\ll S,S\gg &\approx& 4 N_0
\displaystyle\frac{(2t)^2}{\omega^2-\omega_S^2} 
&4 N_0 \displaystyle\frac{\omega_S^2}{\omega^2-\omega_S^2}\,,
\\
\ll \Phi_T,S\gg &=& 0&4 i N_0
\displaystyle\frac{\omega_0^2}{\omega^2-\omega_S^2} 
                   \displaystyle \frac{\omega}{2\Delta} \,
                   \displaystyle \frac{V_\perp -J}{2J} \,,
\\
\ll A_T, S\gg &\approx& 4 N_0\displaystyle
\frac{\omega_0^2}{\omega^2-\omega_S^2} 
                    \displaystyle\frac{t}{\Delta} \,
                   \displaystyle \frac{V_\perp
                     -J}{V_\parallel+V_\perp+2J} 
&0\,, \\
\ll M,S\gg &\approx& - 4i N_0
\displaystyle\frac{2t\omega}{\omega^2-\omega_S^2 }
& 
 - 4i N_0 \displaystyle\frac{2t\omega}{\omega^2-\omega_S^2}\,, 
\\
\omega_S^2 &=& (2t)^2+\omega_0^2&(2t)^2+\omega_0^2
\,, 
\\
\omega_0^2& = & 
\displaystyle\frac{-(V_\parallel-V_\perp+2J)}
{(V_\perp-J)(V_\parallel+J)}
            \displaystyle \frac{(2\Delta)^2}{N_0}
& \displaystyle\frac{-2J}{V_\perp^2-J^2}\frac{(2\Delta)^2}{N_0}\,.
\end{array}
\label{approx}
\end{equation}
\renewcommand{\arraystretch}{1.}
A spin polarization $S$ with 
opposite  sign in the two layers obviously  induces   inter-layer 
triplet-amplitudes $A_T$ ($\Phi_T$).

These results are closely connected to the collective modes
discovered in  
 density-density-correlation functions like $ \ll P, P \gg $ 
($P := -P^{33} 
$)
in our previous work.\cite{wir} 
For pure inter-(intra)-layer pairing one has the exact relation
\begin{equation} \label{intrainter}
\ll P, P \gg_{\rm inter ( intra ) } = \ll S, S \gg_{\rm intra (
  inter ) } , 
\end{equation}
which follows from a unitary change of representation
$\tilde{A} = U A U^{\dagger},\tilde{\mid \psi \rangle} = U 
\mbox{$\mid \psi \rangle$} $
 with 
$U :=
\exp( - i \pi \sum_{k} ( c^{\phantom\dagger}_{1 k \downarrow} c_{2 k 
  \downarrow}^{\dagger}  
+ c^{\phantom\dagger}_{2 k \downarrow} c_{1 k
  \downarrow}^{\dagger})) $

The Goldstone theorem 
\cite{goldstone} predicts the existence of 
excitations with vanishing energy,
 if a continuous, dynamical symmetry $\Omega$
 ($[ \Omega, H ] = 0$) 
is spontaneously broken, e.g.  the  groundstate $ \mid 0 \, \rangle$
is  
not an eigenstate of $\Omega$. Thereby it can help to classify the
resonances  
found in the correlation functions (\ref{correl})  as 
so-called Goldstone modes connected
with certain symmetries of $H$.

Assuming pure singlet pairing in equilibrium, the superconducting 
groundstate is given by
\begin{equation}
\mid \theta_{\parallel} , \theta_{\perp} \, \rangle = 
\prod_{k} \left( 1  + \alpha_{k \parallel} e^{i 2
    \theta_{\parallel}} 
 \Delta_{k \parallel, S}^{\dagger}
 + \alpha_{k \perp} e^{i 2 \theta_{\perp}} \Delta_{k
   \perp,S}^{\dagger} ) 
    \right) 
 \mid 0 \, \rangle
\end{equation}
with $\Delta_{\parallel, S} := \Delta_{11} + \Delta_{22}$ and 
$\Delta_{\perp, S} := \Delta_{12} + \Delta_{21}$ being the 
singlet-order parameters for intra- and inter-layer pairing,
respectively. The analytical calculations of case i (ii) refer to
pure  
intra-(inter-)layer pairing  with
$\alpha_{\perp}=0$ ($\alpha_{\parallel}=0 $).

Table \ref{Tabellespontan} shows (for different parameters $t, J , 
V_{\parallel},  V_{\perp}$) the symmetries $\Omega_{ij}(\phi) := 
\exp (i \phi P^{ij} )$, which are broken in the presence of 
intra-(inter)-layer pairing $\alpha_{\parallel} \neq 0$
($\alpha_{\perp} \neq 0$). 
\begin{table}
\caption{\label{Tabellespontan}  Broken symmetries if
 $\alpha_{\parallel} \neq 0$ 
or $\alpha_{\perp} \neq 0$.
}
\begin{center}
\begin{tabular}{|c|c|c|c|}
\hline
case &parameters& $\alpha_{\parallel} \neq 0$   &  $\alpha_{\perp}
\neq 0$ 
\\
\hline
\hline
1&$t,J, V_{\parallel} - V_{\perp} = 0   $&
$\Omega_{03},\,\,\Omega_{33},\,\,\Omega_{13},\,\,\Omega_{23}$& 
$\Omega_{03},\,\,\Omega_{13},\,\,\Omega_{30},\,\,\Omega_{20}$ 
\\
\hline
2&$t,J\ne0,\, V_{\parallel} =  V_{\perp}$ &
$\Omega_{03},\,\,\Omega_{13}$& 
 $\Omega_{03},\,\,\Omega_{13}$
\\
\hline
3&$t,J\to0,\, V_{\parallel} \neq V_{\perp}  $&
$\Omega_{03},\,\,\Omega_{33}$&  
 $\Omega_{03},\,\,\Omega_{30}$
\\
\hline
4&$t,J,  V_{\parallel} - V_{\perp}  \neq 0$& $\Omega_{03}$& 
 $\Omega_{03}$
\\
\hline
\end{tabular}
\end{center}
\end{table}
The spontaneous breakdown 
$ \Omega_{03} \mid \theta_{\parallel} , \theta_{\perp} \, \rangle =  
\mbox{$\mid \theta_{\parallel} + \phi  , \theta_{\perp} + \phi \,
  \rangle$} $ 
of the global gauge symmetry, which is  generated by the total
particle number, 
 in $\mid \theta_{\parallel} , \theta_{\perp} \, \rangle$
 is a defining property of 
the superconducting phase (case 4), as it is invariably connected with 
non-vanishing Cooper-pair amplitudes ($\langle \Delta_{ij} \rangle
\neq 0 $). 

In Eq. \ref{correl}  Goldstone modes ($\omega_S = 0$) appear in 
case i (ii), if and only if 
$t, J, V_{\parallel} - V_{\perp} = 0$ ($t, J = 0$). We can identify
these  
resonances in case i (ii) with the  modes in table 
\ref{Tabellespontan} in the cases 1 (3), which are  connected 
 with the symmetries $\Omega_{23}$ ($\Omega_{30}$).
The transformations 
\begin{eqnarray}\scriptstyle
\Omega_{23} (\phi) \mid \theta_{\parallel}, \theta_{\perp} \rangle
&=\scriptstyle &\scriptstyle 
\prod_{k} \left( 1  + \alpha_{\parallel} e^{i 2 \theta_{\parallel}} 
\cos (2 \phi) \Delta_{\parallel, S}^{\dagger}
 + \alpha_{\perp} e^{i 2 \theta_{\perp}}
( \Delta_{\perp,S}^{\dagger} -  \sin(2 \phi ) 
\Delta_{\perp, T}^{\dagger} ) \right) 
 \mid 0 \, \rangle, \\
\scriptstyle
\Omega_{30} (\phi) \mid \theta_{\parallel}, \theta_{\perp} \rangle
&=\scriptstyle &\scriptstyle 
\prod_{k} \left( 1  + \alpha_{\parallel} e^{i 2 \theta_{\parallel}}
 \Delta_{\parallel, S}^{\dagger}
 + \alpha_{\perp} e^{i 2 \theta_{\perp}}
(\cos (2 \phi) \Delta_{\perp,S}^{\dagger}  -  i\sin(2 \phi ) 
\Delta_{\perp, T}^{\dagger} ) \right) 
 \mid 0 \, \rangle \nonumber
\end{eqnarray}
show that in both cases the $S_z$=0-component $\Delta_{\perp, T}$ of
the 
triplet-order parameter is excited, which for 
$\Omega_{23}$ ($\Omega_{30}$) also 
creates non-vanishing expectation values  $ \langle A_T \rangle $ 
($\langle \Phi_T \rangle $) and finite responses  
$\ll A_T, S \gg $ ($\ll \Phi_T , S \gg $) to an external spin
polarization  
$S$.  
Thereby $\Omega_{23}$ mixes  intra-layer pairs $\langle
\Delta_{\parallel,S}\rangle$with  triplet-inter-layer pairs $\langle
\Delta_{\perp,T}\rangle$, which is connected with a  spin transfer
between  
the layers without breaking up Cooper pairs. On the other hand, 
$\Omega_{30}$ leaves the modulus of the pairing amplitudes
invariant,  
but creates a phase difference between $\Delta_{12}$ and
$\Delta_{21}$,  
which is the origin of a  supercurrent of inter-layer pairs, the
so-called  
{\em spin Josephson-effect}. 
This terminology is motivated by the close analogy to the usual
Josephson  
effect, where a charge rather than a spin transfer is driven by a
phase  
difference of intra- rather than inter-layer pairs.  The 
 density-modes $\Omega_{20}$ ($\Omega_{33}$), 
which correspond  according to the relation (\ref{intrainter}) to  
the spin modes $\Omega_{23}$ ($\Omega_{30}$),  can be observed  as
poles in  
$\ll P, P \gg$ calculated in our  previous work \cite{wir} rather
than in  
$\ll S, S \gg$.
According to Eq. \ref{correl} all these modes cannot be excited in
the  
absence of particle transfer ($t, J = 0$)
between the layers. 
Finally, $\Omega_{13}$ connects 
groundstates with different ratios of inter- and 
intra-layer-pairing, which  for $t, J = 0$ in case 1 and 2
are  energetically degenerate.

\section{NUMERICAL RESULTS}

We carried out numerical calculations for $\ll S,S\gg$ using  
two slightly different effective masses
$
2m_1  /  \hbar^2 = 1\,\, {\rm eV^{-1}} \AA^{-2},\quad
 2m_2 / \hbar^2 =1.2\,\, {\rm eV^{-1}} \AA^{-2}
$ 
for the two bands $\epsilon_k \pm t_k= \hbar^2  k^2 / 2 m_{2/1}$
and parameters: $\mu = 0.3$  eV,  $\omega_c = 0.25$ eV (cut-off in
k-space), 
$
 ( V_\parallel+V_\perp+ 2 J) N_0  = -0.44, \quad
(V_\parallel-V_\perp) N_0 = \pm 0.2 
$ ($N_0$:  averaged density of states of the two bands).
Then the pairing is mixed and 
for $(V_\parallel-V_\perp)N_0  < 0$ ($ > 0$)  dominated by 
intra-(inter)-layer pairing.

\begin{figure} [h]
\begin{center}
\leavevmode
\epsfxsize=1.\linewidth
\epsfbox{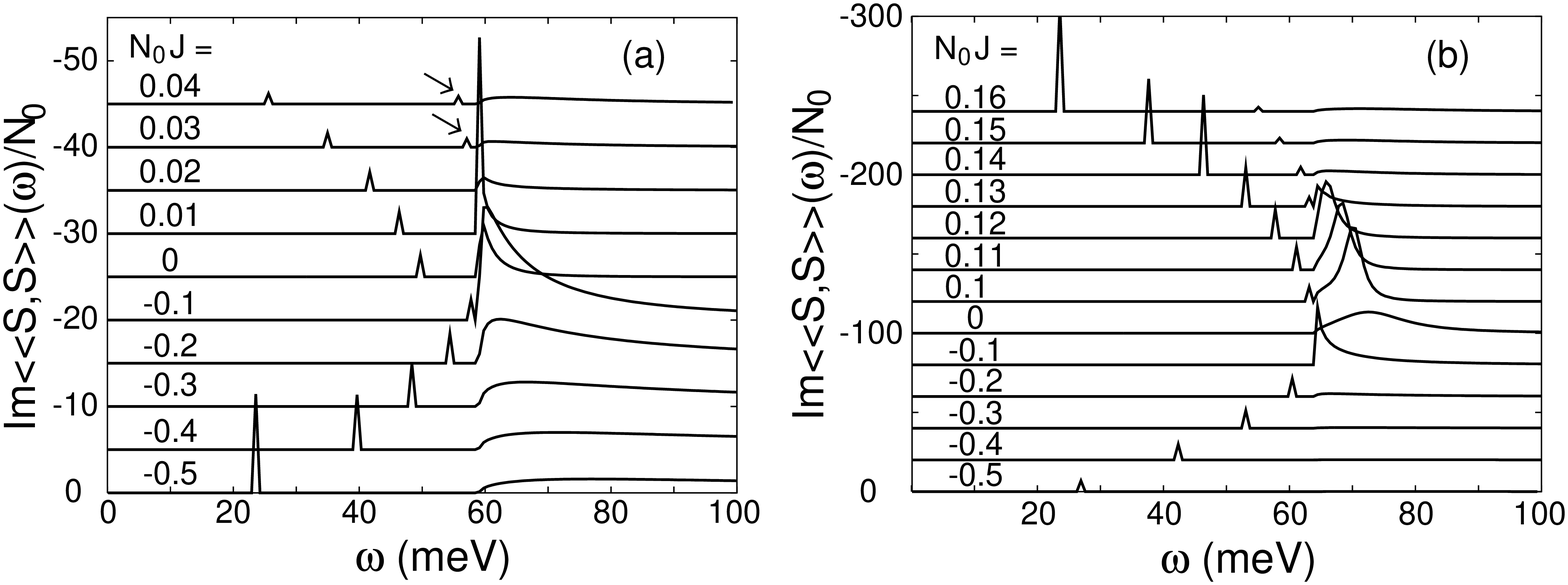}
\caption{\label{SSinterintra}
${\rm Im} \ll S , S \gg $ at $T=0$ in the case of {\em dominant} 
inter-layer (a)  and intra-layer (b) pairing for different $J$.
The spectral weight of the collective mode is given by the height of
the  
$\delta$-peak times 10 in (a) or times 100 for peaks marked with 
arrows and in (b). 
}
\end{center}
\end{figure}
Fig. \ref{SSinterintra}  shows the imaginary part of the
spin-polarization function for different $J$ in the
case a (b) of dominant inter-(intra)-layer pairing at $T=0$.
The collective modes appear as $\delta$-peaks below the onset of
particle hole  
excitations around 60  meV. 
In the appropriate parameter range  the resonances in case a (b) 
coincide with 
the poles  calculated analytically in Eq. 
 \ref{correl} in case ii (i), 
but they exist in a much larger parameter range.
 For larger positive
or negative 
$J$-values than given in the figures the mode frequencies pass
zero indicating an instability of the system.

The spectral weight of the phase mode decreases for dominant
inter-layer pairing going from negative to positive $J$ as indicated
by the spectral weight $\omega_S^2$ in the approximation formula
(\ref{approx}).

For positive $J$ a further collective mode, the so-called 
amplitude mode,  can be seen in both
cases. It is   inside the particle-hole spectrum for small
$J$, but undamped for large $J$ (arrows (a) or small peaks below
the particle-hole threshold (b)).
 Because of the mixing of intra-layer and inter-layer
pairing the spin excitation couples to both the phase  $\Phi_T$ 
and the amplitude $A_T$
of the triplet order-parameter. This causes two collective modes,
which were  
 already  discussed in \cite{wir}. 

\begin{figure} [h]
\begin{center}
\leavevmode
\epsfxsize=1.\linewidth
\epsfbox{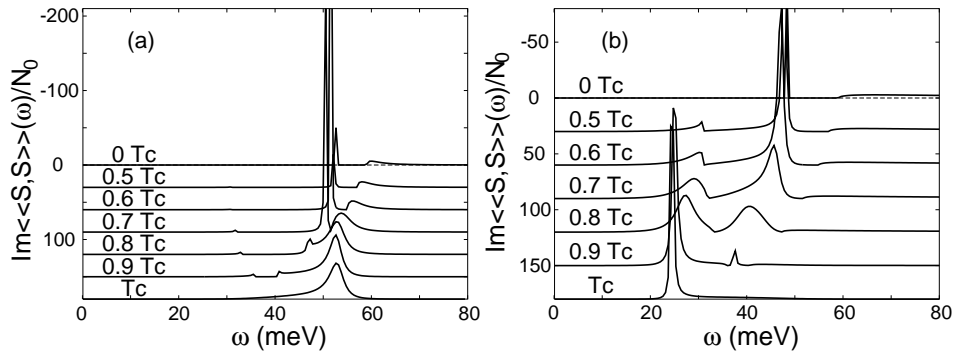}
\caption{\label{SSintertemp} ${\rm Im} \ll S, S \gg $
 for
different temperatures in the case  of dominant  
inter-layer pairing
for constant $N_0 J = 0.01$ or constant  $N_0 J =  -0.3$.}
\end{center}
\end{figure}
The temperature dependence of the spin polarization function in Fig.

\ref{SSintertemp} shows the broadening of the collective mode at
about  
50 meV with increasing $T$. In case b a further collective mode
appears at 
 $T_c$.

To conclude, the anti-ferromagnetic spin polarization couples to the 
phase and amplitude of the triplet-order parameter with
$S_z=0$. This causes a 
collective mode where spin-up and spin-down particle tunnel in
opposite direction ({\em spin Josephson-effect}).  
This might be connected with a resonance 
found in magnetic neutron scattering on Y-Ba-Cu-O at finite $q$,
which shows the same temperature dependence as our mode\cite{Fong}.

This  work  has  been  supported  by  grants  from  the
Studienstiftung des Deutschen Volkes (C.H.) and    the    Bayerische
Forschungsstiftung within the research project FORSUPRA II (F.F.).


\end{document}